\documentclass[secnumarabi,superscriptaddress,amssymb,nobibnotes,aps,prc]{revtex4}

\usepackage{epsfig}
\usepackage{amsmath}
\usepackage{graphicx}
\usepackage{hyperref}

\begin{document}


\title{Inclusive dijet photoproduction in ultraperipheral heavy-ion collisions
 at the LHC in next-to-leading order QCD}

\author{V. Guzey}

\affiliation{Department  of  Physics,  University  of  Jyv\"askyl\"a, P.O.
 Box 35, 40014  University  of  Jyv\"askyl\"a,  Finland}
\affiliation{Helsinki Institute of Physics, P.O.  Box  64,  00014 
 University  of  Helsinki,  Finland}
\affiliation{National Research Center ``Kurchatov Institute'', Petersburg
 Nuclear Physics Institute (PNPI), Gatchina, 188300, Russia}
 
\author{M. Klasen}

\affiliation{Institut f\"ur Theoretische Physik, Westf\"alische
 Wilhelms-Universit\"at M\"unster, Wilhelm-Klemm-Stra{\ss}e 9, 48149
 M\"unster, Germany}




\begin{abstract}

We compute the cross section of inclusive dijet photoproduction in
ultraperipheral Pb-Pb collisions at the LHC using next-to-leading order
perturbative QCD. We demonstrate that our theoretical calculations provide
a good description of various kinematic distributions measured by the ATLAS
collaboration. We find that the calculated dijet photoproduction cross
section is sensitive to nuclear modifications of parton distribution functions
(PDFs) at the level of 10 to 20\%. Hence, this process can be used to reduce
uncertainties in the determination of these nuclear PDFs, whose current
magnitude is comparable to the size of the calculated nuclear modifications
of the dijet photoproduction cross section.

\end{abstract}

\maketitle

\section{Introduction}
\label{sec:intro}

Ultraperipheral collisions (UPCs) of relativistic ions correspond to large
impact parameters between the nuclei exceeding the sum of their radii, so
that short-range strong interactions between the ions are suppressed and
reactions proceed rather via the emission of quasi-real photons by the
colliding ions. Thus, UPCs allow one to study photon-photon and photon-hadron
(proton, nucleus) interactions at high energies
\cite{Baltz:2007kq}.
During the last decade, UPCs have become an active field of research, driven
by experimental results obtained at the Relativistic Heavy Ion Collider (RHIC)
and the Large Hadron Collider (LHC), for a recent experimental review see,
e.g., \cite{Klein:2017vua}. Notable examples of various UPC processes and
their analyses include the two-photon production of dilepton pairs
\cite{Klein:2018cjh,Goncalves:2018gca}; light-by-light scattering $\gamma
\gamma \to \gamma \gamma$ and searches for potential physics beyond the
Standard Model \cite{dEnterria:2013zqi,Fichet:2014uka,Klusek-Gawenda:2016euz};
an electromagnetic double-scattering contribution to dimuon pair production
in photon-photon scattering \cite{vanHameren:2017krz}; exclusive
photoproduction of charmonia in proton-proton \cite{Aaij:2013jxj,Aaij:2014iea},
proton-nucleus \cite{TheALICE:2014dwa} and nucleus-nucleus
\cite{Abbas:2013oua,Abelev:2012ba,Adam:2015sia,Khachatryan:2016qhq} UPCs and
of bottomonia in proton-proton \cite{Aaij:2015kea} and proton-nucleus UPCs
\cite{Chudasama:2016eck}; new constraints on the small-$x$ gluon distribution
in the proton \cite{Jones:2013pga,Guzey:2013qza} and heavy nuclei
\cite{Adeluyi:2012ph,Guzey:2013xba} and the dynamics of strong interactions at
high energies in the color dipole framework \cite{Lappi:2013am,%
Goncalves:2014wna,Mantysaari:2017dwh}; and exclusive photoproduction of $\rho$
mesons on nuclei \cite{Adler:2002sc,Abelev:2007nb,Agakishiev:2011me,%
Adam:2015gsa} as well as tests of models of nuclear shadowing
\cite{Rebyakova:2011vf,Frankfurt:2015cwa}.

Focusing on UPC studies of nuclear structure in QCD at the LHC, coherent
$J/\psi$ photoproduction in Pb-Pb UPCs at $\sqrt{s_{NN}} = 2.76$ TeV
\cite{Abbas:2013oua,Abelev:2012ba,Adam:2015sia,Khachatryan:2016qhq} revealed
a significant nuclear suppression of the measured rapidity distributions. In
the framework of the leading logarithmic approximation of perturbative QCD
\cite{Ryskin:1992ui}, it can be interpreted as evidence of large nuclear
gluon shadowing, $R_g=f_{g/A}(x,\mu^2)/[Af_{g/N}(x,\mu^2)] \approx 0.6$ at $x=
10^{-3}$ and $\mu^2=3$ GeV$^2$ ($f_{g/A}$ and $f_{g/N}$ are gluon densities in
Pb and the proton, respectively). This value of $R_g$ agrees with predictions
of the leading twist nuclear shadowing model \cite{Frankfurt:2011cs}, which
are characterized by small theoretical uncertainties in this kinematic region.
It is also broadly consistent with the EPS09 \cite{Eskola:2009uj}, nCTEQ15
\cite{Kovarik:2015cma}, and EPPS16 \cite{Eskola:2016oht} nuclear parton
distribution functions (nPDFs), which however have significant uncertainties
in this kinematic regime. Note that in the collinear factorization framework,
next-to-leading order (NLO) perturbative QCD corrections to the cross section
of $J/\psi$ photoproduction are large \cite{Ivanov:2004vd,Jones:2015nna} and
the relation between the gluon parton distribution function (PDF) and the
gluon generalized parton distribution (GPD) is model-dependent, which makes it
challenging to interpret the UPC data on $J/\psi$ photoproduction on nuclei in
terms of the NLO gluon nPDF.

The program of UPC measurements continues with Run 2 at the LHC, where besides
photoproduction of vector mesons, inclusive dijet photoproduction in Pb-Pb
UPCs $AA \to A+{\rm 2 jets}+X$ has also recently been measured by the ATLAS
collaboration \cite{Atlas} (for leading-order QCD predictions for rates of
this process, see \cite{Strikman:2005yv}). The cross section of this process
is sensitive to quark and gluon nPDFs $f_{j/A}(x,\mu^2)$ in a wide range of
the momentum fraction $x$ and the resolution scale $\mu > {\cal O}(20)$ GeV,
where one still expects sizable nuclear modifications of the PDFs. In addition,
imposing the requirement that the target nucleus stays intact, one can study
diffractive dijet photoproduction in UPCs $AA \to A+{\rm 2 jets}+X+A$. Studies
of this process may shed some light on the mechanism of QCD factorization
breaking in diffractive photoproduction and, for the first time, give access
to nuclear diffractive PDFs \cite{Guzey:2016tek,Basso:2017mue}. While further progress in
constraining nPDFs will benefit from studies of high-energy hard processes
with nuclei in proton-nucleus ($pA$) scattering at the LHC
\cite{Salgado:2011wc} and lepton-nucleus ($eA$) scattering at a future
Electron-Ion Collider (EIC) \cite{Accardi:2012qut} and LHeC
\cite{AbelleiraFernandez:2012cc}, UPCs at the LHC present an important and
complimentary method of obtaining new constraints on nPDFs in a wide
kinematic range already now.

In this work, we make predictions for the cross section of inclusive dijet
photoproduction in Pb-Pb UPCs at the LHC using NLO perturbative QCD
\cite{Klasen:2002xb} and nCTEQ15 nPDFs. We show that our approach provides a
good description 
of various cross section distributions measured
by the ATLAS collaboration \cite{Atlas}. 
Our analysis also shows that the dijet photoproduction cross
section in the considered kinematics is sensitive to nuclear modifications of
the PDFs. As a function of the momentum fraction $x_A$, the ratio of the cross
sections calculated with nPDFs and in the impulse approximation behaves
similarly to $R_g$ for a given $\mu$ and deviates from
unity by $10-20$\% for the central nCTEQ15 fit. The calculations using EPPS16
nPDFs and predictions of the leading twist nuclear shadowing model give
similar results. This suggests that inclusive dijet photoproduction on nuclei
can be used to reduce uncertainties in the determination of nPDFs, which are
currently significant and comparable in size to the magnitude of the
calculated nuclear modifications of the dijet photoproduction cross section.

The remainder of this paper is structured as follows. In Sec.
\ref{sec:formalism}, we outline the formalism of dijet photoproduction
in UPCs using NLO perturbative QCD. We present and discuss our results for
the LHC in Sec.\ \ref{sec:results} and draw conclusions in Sec.\
\ref{sec:conclusions}.

\section{Photoproduction of dijets in UPCs in NLO perturbative QCD}
\label{sec:formalism}

\begin{figure}[t]
\begin{center}
\epsfig{file=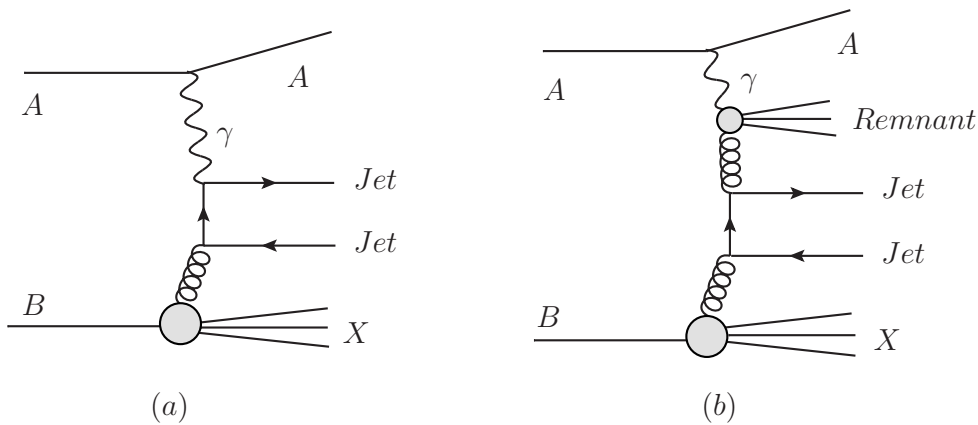,scale=0.8}
\caption{Typical leading-order Feynman graphs for dijet photoproduction in UPCs
 of hadrons $A$ and $B$. Graphs (a) and (b) correspond to the direct and
 resolved photon contributions, respectively.}
\label{fig:photo_jets_upc}
\end{center}
\end{figure}

Typical leading-order (LO) Feynman diagrams for dijet photoproduction in UPCs
of nuclei $A$ and $B$ are shown in Fig.\ \ref{fig:photo_jets_upc}, where the
graphs ($a$) and ($b$) correspond to the direct and resolved photon
contributions, respectively. Note that beyond LO, the separation of the direct
and resolved photon contributions depends on the factorization scheme and
scale 
(see the discussion below).

Using the Weizs\"acker-Williams method, which allows one to treat the electromagnetic field of an ultra-relativistic
ion as a flux of equivalent quasi-real photons~\cite{Baltz:2007kq,Bertulani:2005ru}, and 
the collinear factorization framework for photon-nucleus scattering, 
the cross section of the UPC
process $AB\to A+{\rm 2 jets}+X$ is given by \cite{Klasen:2002xb}
\begin{eqnarray}
&& d\sigma(AB \to A+{\rm 2 jets}+X) = \nonumber\\
&& \sum_{a,b} \int^{y_{\rm max}}_{y_{\rm min}} dy \int^1_0 dx_{\gamma} \int^{x_{A,{\rm max}}}_{x_{A,{\rm min}}} dx_A f_{\gamma/A}(y)f_{a/\gamma}(x_{\gamma},\mu^2)f_{b/B}(x_A,\mu^2) d\hat{\sigma}(ab \to {\rm jets})\,,
\label{eq:cs}
\end{eqnarray}
where $a,b$ are parton flavors; $f_{\gamma/A}(y)$ is the flux of equivalent photons
emitted by ion $A$, which depends on the photon light-cone momentum fraction $y$;
$f_{a/\gamma}(x_{\gamma},\mu^2)$ is the PDF of the photon, which depends on the
momentum fraction $x_{\gamma}$ and the factorization scale $\mu$; $f_{b/B}(x_A,\mu^2)$
is the nuclear PDF with $x_A$ being the corresponding parton momentum fraction;
and $d\hat{\sigma}(ab \to {\rm jets})$ is the elementary cross section for production
of two- and three-parton final states emerging as jets in hard scattering of partons
$a$ and $b$. The sum over $a$ involves quarks and gluons for the resolved photon contribution 
and the photon for the direct photon contribution dominating at $x_{\gamma} \approx 1$.
At LO, the direct photon contribution has support exactly only at $x_{\gamma} = 1$,
i. e., $f_{a_/\gamma}=\delta(1-x_\gamma)$.
At NLO, the virtual and real corrections are calculated with massless
quarks in dimensional regularization, ultraviolet (UV) divergences are renormalized in
the $\overline{\rm MS}$ scheme, infrared (IR) divergences are canceled and factorized
into the proton and photon PDFs, respectively. For the latter, this implies
a transformation from the DIS$_\gamma$ into the $\overline{\rm MS}$
scheme.
The integration limits are determined by the rapidities and transverse momenta
 of the produced jets, see Sec.~\ref{sec:results}.
Note that Eq.~(\ref{eq:cs}) is based on 
the clear separation of scales, which characterize the long-distance electromagnetic interaction
and the short-distance strong interaction. 
It generalizes  
the NLO perturbative QCD formalism 
of collinear factorization for jet photoproduction in lepton-proton 
scattering developed in Refs.~\cite{Klasen:1995ab,Klasen:1996it,Klasen:1997br,Klasen:2002xb}, which
successfully described HERA $ep$ data on dijet photoproduction~\cite{Klasen:2011ax}.
Hence, Eq.~(\ref{eq:cs}) involves universal nuclear PDFs $f_{b/B}(x_A,\mu^2)$, which can be accessed 
in a variety of hard processes involving nuclear targets~\cite{Eskola:2009uj,Kovarik:2015cma,Eskola:2016oht}, 
and the universal photon PDFs $f_{a/\gamma}(x_{\gamma},\mu^2)$, which are determined by $e^{+}e^{-}$ data, for a review, see \cite{Klasen:2002xb}.
Hence, the interplay between the direct and resolved photon contributions in Eq.~(\ref{eq:cs}) is also universal
and controlled by the standard $\mu^2$ evolution equations of photon PDFs and the choice of the factorization scheme.

In our analysis, we used the following input for Eq.~(\ref{eq:cs}).
For photon PDFs $f_{a/\gamma}(x_{\gamma},\mu^2)$, we used 
the GRV HO parametrization~\cite{Gluck:1991jc}, which we transformed from the DIS$_{\gamma}$ to the $\overline{{\rm MS}}$ factorization scheme. These photon PDFs have been tested profoundly at HERA and the Large Electron-Positron (LEP) collider at CERN and are very robust,
in particular at high $x_{\gamma}$ (dominated by the pQCD photon-quark splitting), which is correlated with the low-$x_A$ gluons 
and sea quarks in Pb that present one of the points of interest of the present study.
For nuclear PDFs $f_{b/B}(x_A,\mu^2)$, we employed the nCTEQ15 parametrization~\cite{Kovarik:2015cma}. 
The photon flux $f_{\gamma/A}(y)$ produced by a relativistic point-like charge $Z$ is given by 
 the standard expression
 \begin{equation}
 f_{\gamma/A}(y)=\frac{2 \alpha_{\rm e.m.}Z^2}{\pi}\frac{1}{y} \left[\zeta K_0(\zeta)K_1(\zeta)-\frac{\zeta^2}{2}(K_1^2(\zeta)-K_0^2(\zeta)) \right] \,,
 \label{eq:flux}
 \end{equation}
where $\alpha_{\rm e.m.}$ is the fine-structure constant; $K_{0,1}$ are modified Bessel functions of the second kind;
$\zeta=y m_p b_{\rm min}$ with $m_p$ being the proton mass and $b_{\rm min}$ the minimal distance between two nuclei.
For Pb-Pb UPCs, Eq.~(\ref{eq:flux}) with $b_{\rm min}=14.2$ fm reproduces very well the photon flux calculated taking into account the nuclear form factor and the suppression of strong interactions at impact parameters $b < b_{\rm min}$, see the discussion in~\cite{Nystrand:2004vn}.

The NLO calculation of the dijet photoproduction cross section using Eq.~(\ref{eq:cs}) is numerically implemented in an NLO 
parton-level Monte Carlo~\cite{Klasen:1995ab,Klasen:1996it,Klasen:1997br,Klasen:2002xb}, which has been successfully tested in many
different environments (HERA, LEP, Tevatron). It implements the anti-k$_T$ algorithm (but we have at most two partons in the jet)
and all the kinematic conditions and cuts used in the ATLAS analysis~\cite{Atlas} that are explicitly explained in the 
following section. Hadronization corrections and underlying event (UE) subtractions are not part of our analysis, but they are expected
to be performed with PYTHIA simulations by the experiment once the data are final (as has been done at HERA).

\section{Predictions for dijet photoproduction in Pb-Pb collisions at the LHC}
\label{sec:results}

The main goal of the present paper is the first NLO QCD calculation of the cross section of inclusive dijet photoproduction 
in Pb-Pb UPCs and the conclusion whether it can describe the results of the ATLAS measurement~\cite{Atlas}.
The ATLAS analysis was performed using
the following conditions and selection criteria: 
\begin{itemize}
\item
the anti-k$_T$ algorithm with the jet radius $R=0.4$;
\item
  the leading jet has $p_{T,1} > 20$ GeV, while the other jets have a different cut on
  $p_{T,i\neq1}> 15$ GeV as
required \cite{Klasen:1995xe}, which corresponds to $35 < H_T < 400$ GeV, where $H_T=\sum_ip_{T,i}$;
\item
all jets have rapidities $|\eta_i| < 4.4$;
\item
the combined mass of all reconstructed jets is $ 35 < m_{\rm jets} < 400$ GeV;
\item
the parton momentum fraction on the photon side $z_{\gamma}=y x_{\gamma}$, $10^{-4} < z_{\gamma} < 0.05$;
\item
the parton momentum fraction on the nucleus side $x_A$, $5 \times 10^{-4} < x_A < 1$.
\end{itemize}

The ATLAS results are presented as distributions in terms of the total jet transverse momentum $H_T=\sum_ip_{T,i}$
and the photon $z_{\gamma}$ and nucleus $x_A$ light-cone momentum fractions
\begin{equation}
z_{\gamma}=\frac{m_{\rm jets}}{\sqrt{s_{NN}}}e^{y_{\rm jets}} \quad , \quad
x_{A}=\frac{m_{\rm jets}}{\sqrt{s_{NN}}}e^{-y_{\rm jets}} \,,
\label{eq:estimators}
\end{equation}
where
\begin{equation}
m_{\rm jets}=\left[\left(\sum_i E_i\right)^2-\left|\sum_i \vec{p}_i \right|^2 \right]^{1/2} \quad , \quad
y_{\rm jets}=\frac{1}{2} \ln \left(\frac{\sum_i E_i+p_{z,i}}{\sum_i E_i-p_{z,i}} \right) \,.
\label{eq:estimators2}
 \end{equation}
 In Eqs.~(\ref{eq:estimators2}), the index $i$ runs over all measured jets; $E_i$ and $\vec{p}_i$ denote the jet energy and momentum, respectively.
 Note that at LO, the kinematics of $2 \to 2$ parton scattering and the momentum fractions $z_{\gamma}$ and $x_A$
  can be exactly reconstructed from the dijet measurement.
 At NLO, Eqs.~(\ref{eq:estimators}) serve as hadron-level estimators of the momentum fractions entering Eq.~(\ref{eq:cs});
 for brevity, we use the same notations in Eqs.\ (\ref{eq:cs}) and (\ref{eq:estimators2}).

\begin{figure}[t]
\begin{center}
\epsfig{file=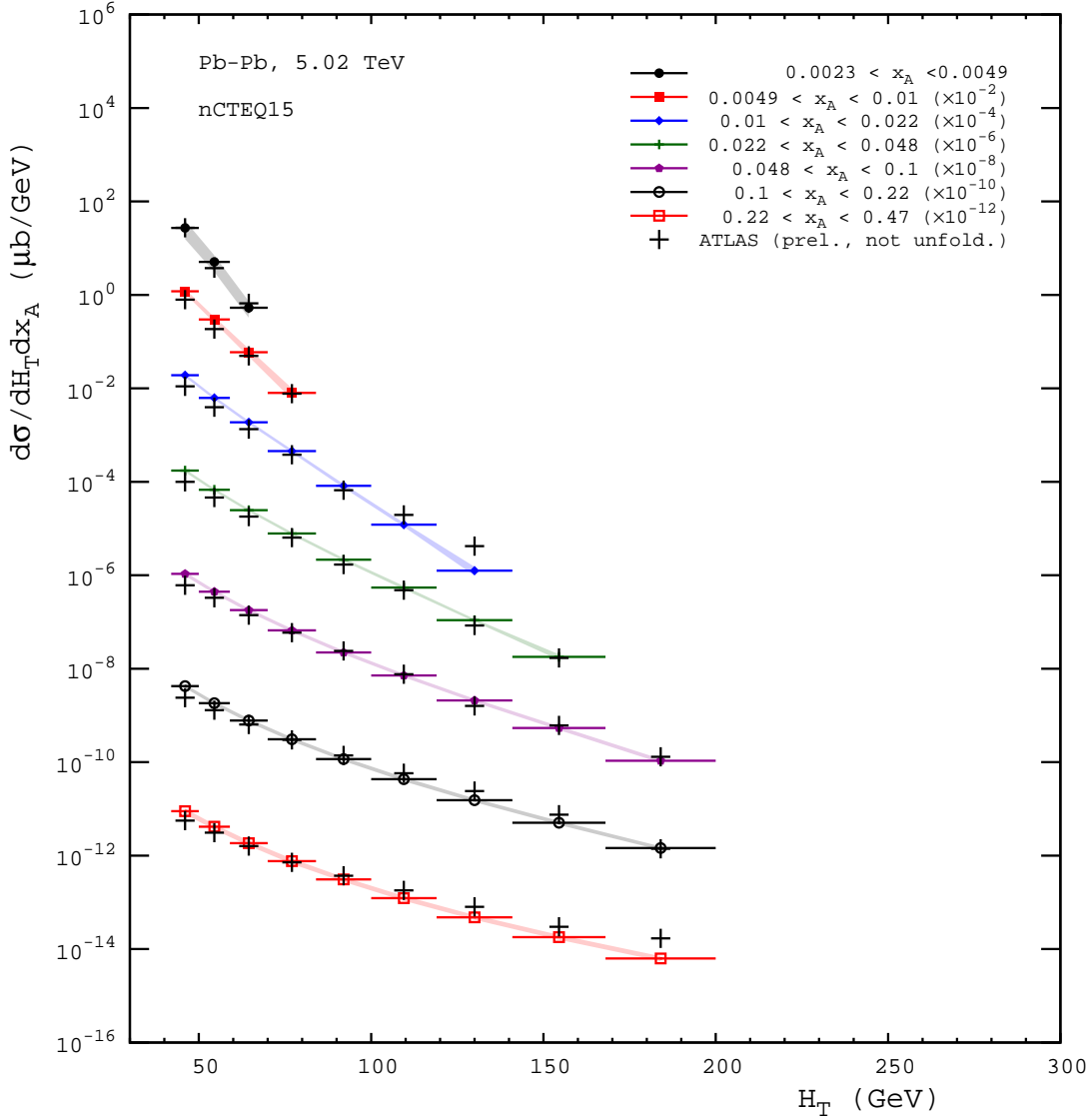,scale=1.15}
\caption{NLO QCD predictions for the cross section of dijet photoproduction in Pb-Pb UPCs at $\sqrt{s_{NN}}=5.02$ TeV in the ATLAS kinematics as a function of $H_T$ for different bins of $x_A$. The central values and the corresponding shaded 
uncertainty bands are obtained using nCTEQ15 nPDFs.
The crosses are the ATLAS data points that we extracted from \cite{Atlas}.}
\label{fig:HT}
\end{center}
\end{figure}
 
\begin{figure}[t]
\begin{center}
\epsfig{file=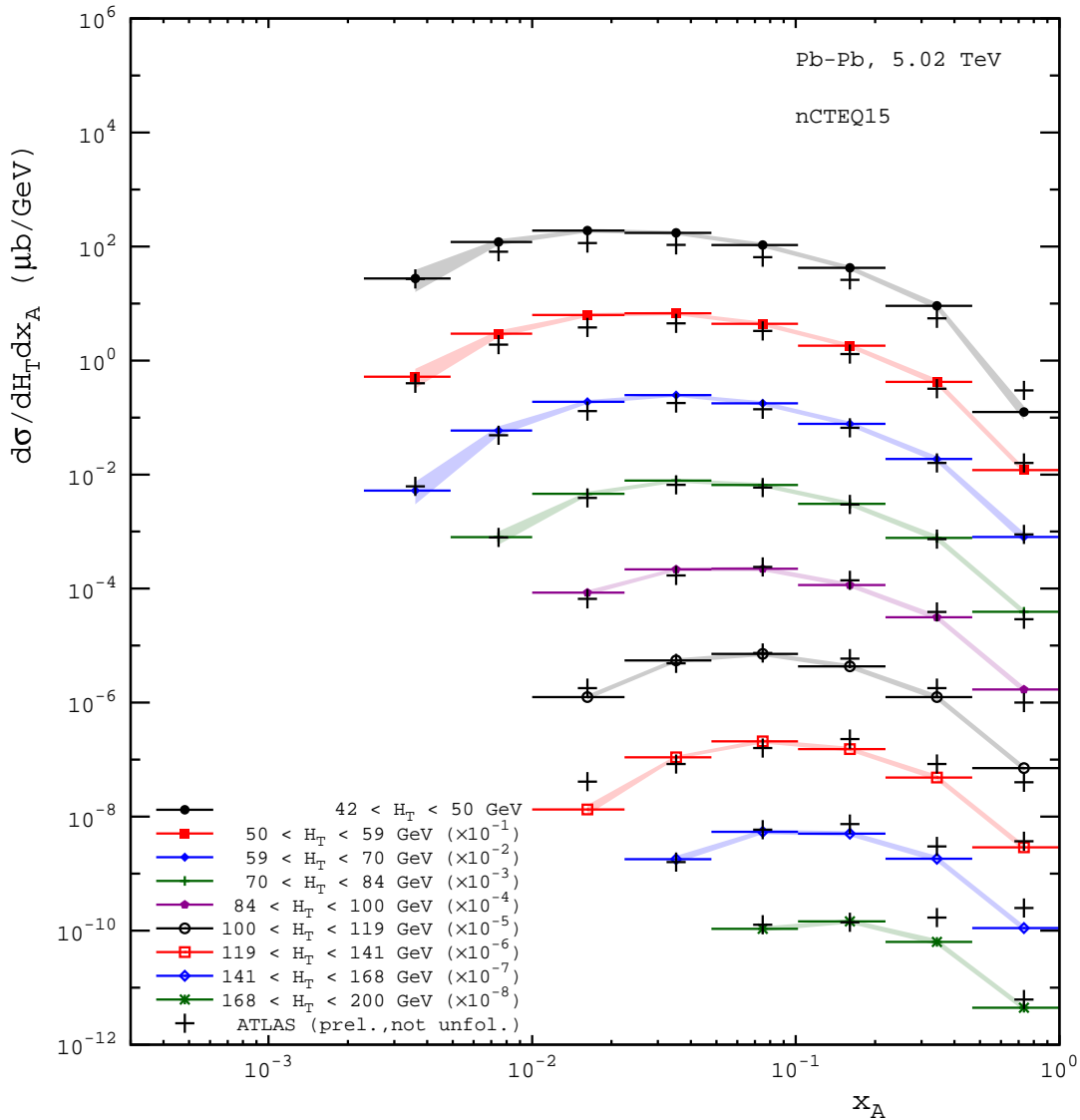,scale=1.15}
\caption{NLO QCD predictions for the cross section of dijet photoproduction in Pb-Pb UPCs at $\sqrt{s_{NN}}=5.02$ TeV in the ATLAS kinematics as a function of $x_A$ for different bins of $H_T$.
The crosses are the ATLAS data points that we extracted from \cite{Atlas}. }
\label{fig:xA}
\end{center}
\end{figure}

\begin{figure}[t]
\begin{center}
\epsfig{file=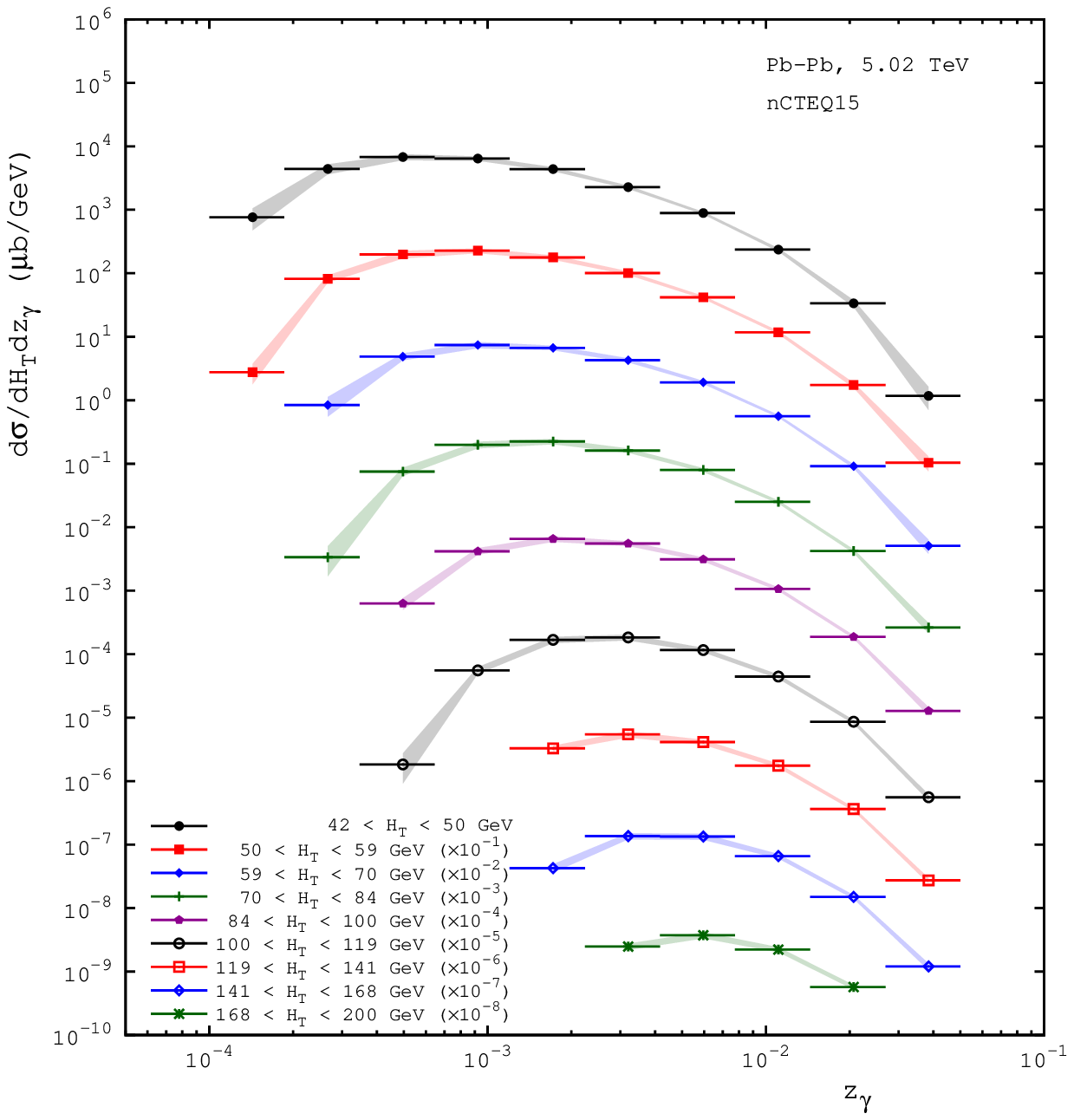,scale=1.15}
\caption{NLO QCD predictions for the cross section of dijet photoproduction in Pb-Pb UPCs at $\sqrt{s_{NN}}=5.02$ TeV in the ATLAS kinematics as a function of $z_{\gamma}$ for different bins of $H_T$.}
\label{fig:zgamma}
\end{center}
\end{figure}

\begin{figure}[t]
\begin{center}
\epsfig{file=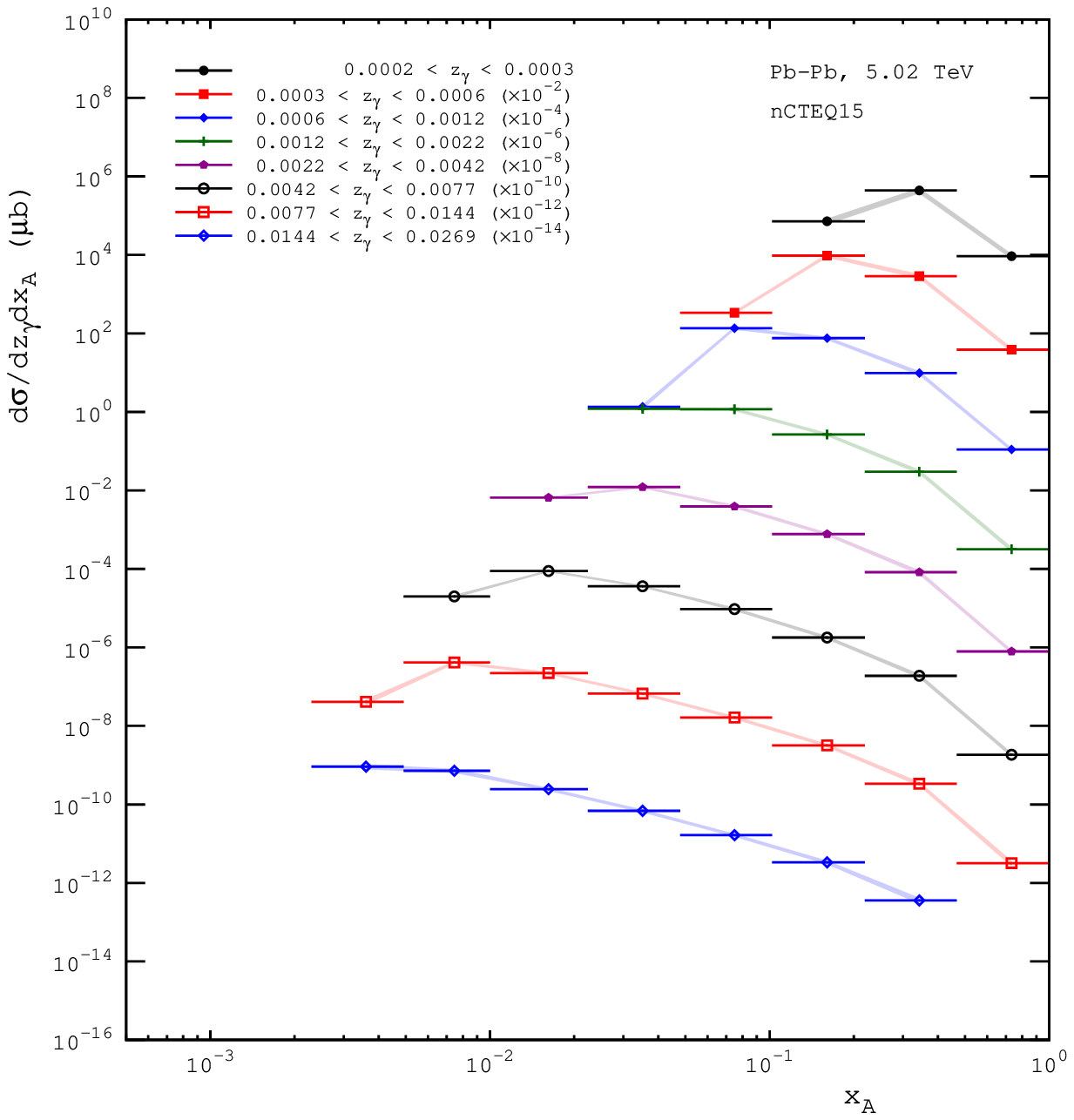,scale=1.15}
\caption{NLO QCD predictions for the cross section of dijet photoproduction in Pb-Pb UPCs at $\sqrt{s_{NN}}=5.02$ TeV in the ATLAS kinematics as a function of $x_A$ for different bins of $z_{\gamma}$.}
\label{fig:xA2}
\end{center}
\end{figure}

Figures~\ref{fig:HT}, \ref{fig:xA}, \ref{fig:zgamma}, and \ref{fig:xA2} show our results for the cross section of dijet photoproduction in Pb-Pb UPCs 
in the ATLAS kinematics (see above) as a function of $H_T$, $x_A$, and $z_{\gamma}$
for different bins of these variables. They correspond to Figs.\ 12-15 of Ref.\ \cite{Atlas}.
In each bin, our predictions are obtained using the central fit of nCTEQ15 nPDFs~\cite{Kovarik:2015cma}. The shaded bands 
quantify the uncertainty of our results $\Delta \sigma$ due to the uncertainty of nCTEQ15 nPDFs. It is calculated by adding in quadrature the individual uncertainties corresponding to each of 32 error sets
\begin{equation}
\Delta \sigma =\frac{1}{2} \sqrt{ \sum_{k=1,\ {\rm odd}}^{31} \left(\sigma(f_k)-\sigma(f_{k+1})\right)^2} \,,
\label{eq:error}
\end{equation}
where $\sigma(f_k)$ is the cross section calculated using the $f_k$ nCTEQ15 error nPDFs.

A comparison of our results shown in Figs.~\ref{fig:HT}, \ref{fig:xA}, \ref{fig:zgamma}, and \ref{fig:xA2}
with Figs.~12-15 of Ref.\ \cite{Atlas} demonstrates that our calculations describe well 
the corresponding distributions.
To illustrate this point, in Figs.~\ref{fig:HT} and \ref{fig:xA}, which present the phenomenologically important 
distributions in $H_T$ and $x_A$, 
respectively, we also explicitly show by crosses the ATLAS data points, which we extracted from \cite{Atlas} using the WebPlotDigitizer
tool~\cite{Digitizer}. One can readily see from these figures that the results of our calculations describe both the shape and 
normalization of the data rather well. The description of the remaining two distributions is also adequate.
Note that the ATLAS data is preliminary and has not been corrected (unfolded) for the detector response.

\begin{figure}[t]
\begin{center}
\epsfig{file=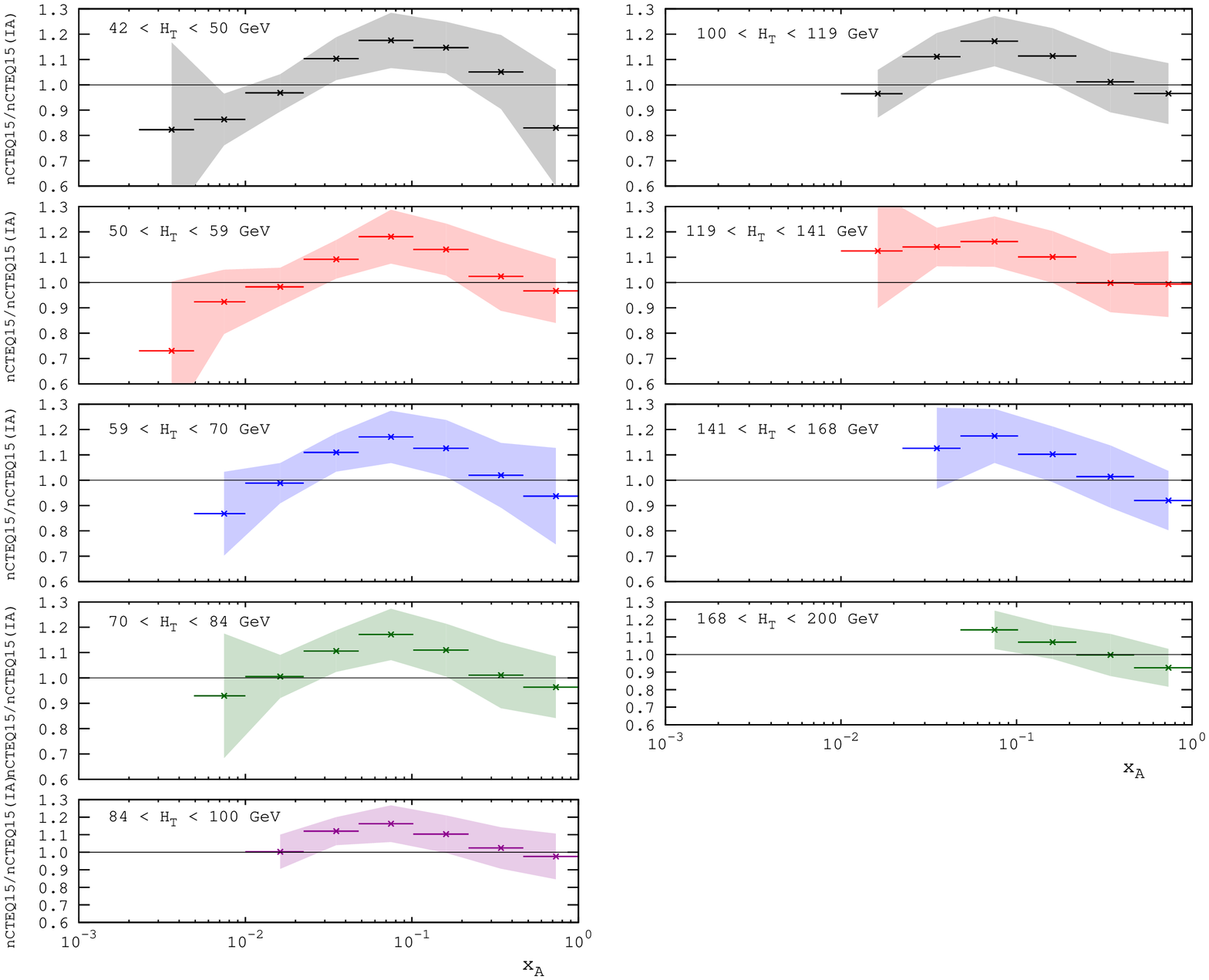,scale=0.7}
\caption{NLO QCD predictions for the ratio of the cross section of dijet photoproduction 
to that calculated in IA in Pb-Pb UPCs at $\sqrt{s_{NN}}=5.02$ TeV in the ATLAS kinematics
as a function of $x_A$ for different $H_T$ bins.
 The shaded bands give the uncertainty of nCTEQ15 nPDFs. }
\label{fig:xA_dopol}
\end{center}
\end{figure}

\begin{figure}[t]
\begin{center}
\epsfig{file=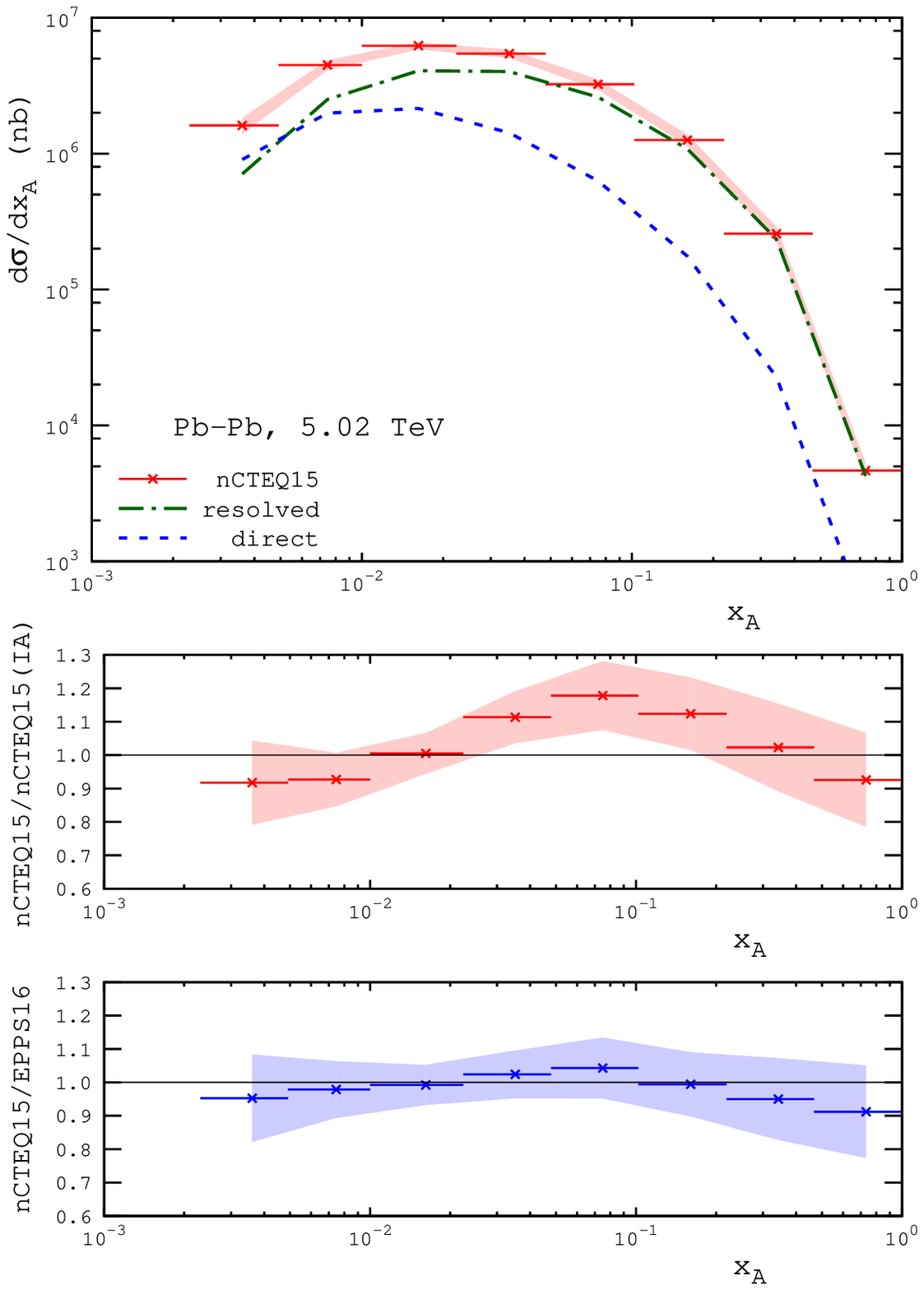,scale=1.0}
\caption{NLO QCD predictions for the cross section of dijet photoproduction in Pb-Pb UPCs at $\sqrt{s_{NN}}=5.02$ TeV in the ATLAS kinematics as a function of $x_A$. Top: The resolved (green, dot-dashed) and direct
(blue, dashed) photon contributions and their sum (red, solid). Middle: The ratio to the impulse approximation. Bottom: The ratio of cross sections calculated using the nCTEQ15 and EPPS16 nPDFs.
The shaded bands show the uncertainty of nCTEQ15 nPDFs.}
\label{fig:xA3}
\end{center}
\end{figure}

To assess the impact of measurements of the inclusive dijet photoproduction cross section on nPDFs,  
we focus on the $x_A$ distribution.
Our results are shown in Figs.~\ref{fig:xA_dopol} and \ref{fig:xA3}. In Fig.~\ref{fig:xA_dopol}, we show 
the ratio of the cross section calculated using nCTEQ15 nPDFs in lead to the 
one calculated in the impulse approximation (IA), where nuclear PDFs are assumed not to include any nuclear modifications and are
given by the weighted sum of free proton and neutron PDFs, 
$f_{b/A}^{\rm IA}=Zf_{b/p}+(A-Z)f_{b/n}$. 
The panels in this figure corresponds to nine bins in $H_T$ presented in Fig.~\ref{fig:xA} (the numerator of the presented
ratio is given by the curves in Fig.~\ref{fig:xA}).
One can see that the cross section ratio as a function of $x_A$ 
behaves similarly to the ratio $R_g = f_{g/A}(x,\mu^2)/[Af_{g/N}(x,\mu^2)]$ of the nuclear and nucleon gluon distributions.
It dips below unity for $x_A < 0.01$ due to nuclear shadowing and then becomes enhanced around $x_A=0.1$ due to the assumed gluon
antishadowing. For $x_A > 0.3$, the cross section ratio shows again a suppression due to
the EMC effect encoded in the nPDFs. 
While the scaling violations (the $H_T$ dependence) of the shown ratios are difficult to see, they seem to be positive
at small $x \sim 0.01$ and negative at large $x$, as they should
be, cf.~\cite{Kovarik:2015cma,Eskola:2016oht}.

Note that in spite of large values of the resolution scale probed in the considered kinematics, $\mu > {\cal O}(20)$ GeV,
one can see that one is still sensitive to nuclear modifications of the PDFs at the $10-20$\% level for the central value of our predictions.
One should also note that the uncertainty due to nPDFs, which is given by the shaded bands, 
is significant and comparable to the size of the discussed nuclear modifications. 
This can be viewed as an opportunity to reduce uncertainties of nPDFs using data on cross section of inclusive dijet photoproduction in nuclei in global QCD fits of nPDFs.

In Fig.~\ref{fig:xA3}, we show our predictions for the dijet cross section as a function of $x_A$ 
integrated over $H_T$ and $z_{\gamma}$.
The top panel presents separately the resolved (green, dot-dashed) and the direct
(blue, dashed) photon contributions to the cross section as well as their sum (red, solid).
As can be expected, because of the 
correlation
between $x_A$ and $z_{\gamma}$, see Eq.~(\ref{eq:estimators}), the resolved photon contribution dominates for $x_A > 0.01$. We find that for small $x_A < 0.01$, the two contributions are comparable with 
the direct contribution being somewhat larger.
While this behavior is qualitatively similar to the results of the LO analysis in the framework of PYTHIA 8 with EPPS16 nPDFs~\cite{Helenius:2018bai}, the relative contribution of the resolved photon term is larger at NLO, but this statement depends of course on the choice of the photon factorization scheme and scale.

The middle panel of Fig.~\ref{fig:xA3} presents the ratio to the impulse approximation.
Similarly to the trend already observed in Fig.~\ref{fig:xA_dopol} and discussed above, 
this ratio as a function of $x_A$ behaves similarly to $R_g$.
This behavior is similar to the one observed in the case of dijet photoproduction in the kinematics of an Electron-Ion Collider 
(EIC)~\cite{Klasen:2018gtb}.

Note that in the future and if/when experimental data become available, to study nuclear modifications of nPDFs one can directly 
form the ratio of dijet cross sections measured in Pb-Pb and proton-proton ($pp$)
UPCs as a function of $x_A$. While the systematics are highly correlated between bins in $x_A$, the information on nPDFs is
in the shape of the cross section ratio, see  the middle panel of Fig.~\ref{fig:xA3} and its discussion above.
Also, while the central Pb-Pb and $pp$ collisions are very different, Pb-Pb and $pp$ UPCs have comparable multiplicities.
Thus, one can expect that the systematic uncertainties largely cancel in the nucleus-to-proton cross section ratio.

Finally, the bottom panel of Fig.~\ref{fig:xA3} presents the ratio of the dijet cross section
calculated using nCTEQ15 nPDFs to the one calculated with the central value of EPPS16 nPDFs. The shaded band
quantifies the uncertainty of the nCTEQ15 fit. One can see from the panel that the two parameterizations
of nPDFs give similar predictions, which differ by at most 5\% for all but one values of $x_A$.
We have also explicitly checked that the use of nPDFs calculated in the model of leading twist nuclear
shadowing~\cite{Frankfurt:2011cs} gives similarly close predictions for the dijet photoproduction cross section.

In our calculations, following the standard prescription for setting the hard scale in QCD calculations, we used 
$\mu = 2 E_{T,1}$ in Eq.~(\ref{eq:cs}). In detail, we performed calculations 
using $\mu = (E_{T,1}/4, E_{T,1}/2, E_{T,1}, 2\,E_{T,1},4\,E_{T,1})$ both at NLO and LO and found that (i) the integrated cross section of inclusive dijet photoproduction at NLO as a function of $\mu$ is approximately 
constant is the vicinity of $\mu = 2 E_{T,1}$, (ii) while the NLO cross section slightly increases with an increase of $\mu$ up to $2E_{T,1}$ and then starts to decrease again,
the LO cross section steeply decreases monotonically, and (iii) the values of the two cross sections are close around $\mu = 2 E_{T,1}$.
Therefore, $\mu = 2 E_{T,1}$ in Eq.~(\ref{eq:cs}) corresponds to the choice, which is most numerically stable
against higher-order corrections.

In this work, we used the framework of collinear factorization and NLO perturbative QCD 
to examine the sensitivity of the dijet photoproduction cross section to nuclear modifications of PDFs. 
Alternatively, one can use this process to look for signs of the 
BFKL and gluon saturation dynamics in the high-energy ($k_{T}$) factorization approach \cite{Kotko:2017oxg}.

\section{Conclusions}
\label{sec:conclusions}

In this work, we calculated the cross section of inclusive dijet photoproduction in Pb-Pb UPCs at the LHC 
using NLO perturbative QCD and nCTEQ15 nPDFs.
We showed that our approach provides a good description of various cross section distributions measured by the ATLAS collaboration.
We found that the calculated dijet photoproduction cross section is sensitive to nuclear modifications of the PDFs. In particular, 
as a function of the nucleus momentum fraction $x_A$, 
the ratio of the cross sections calculated with nPDFs and in the impulse approximation
 behaves similarly to $R_g$ for given $\mu$ and deviates from unity by $10-20$\% for the central nCTEQ15 fit. 
The calculations using EPPS16 nPDFs and predictions of the leading 
twist nuclear shadowing model give similar results.
Therefore, inclusive dijet photoproduction on nuclei has the potential to reduce 
uncertainties in determination of nPDFs, which are comparable to the magnitude 
of the calculated nuclear modifications of the dijet photoproduction cross section.
Our present analysis is a step in this direction.

\acknowledgments

VG would like to thank A.~Angerami, I.~Helenius, K.~Kovarik, H.~Paukkunen, and M.~Strikman  for useful discussions
and the Institut f\"ur Theoretische Physik, Westf\"alische Wilhelms-Universit\"at M\"unster for hospitality. 
VG's research is supported in part by 
RFBR, research project 17-52-12070. The authors gratefully acknowledge financial 
support of DFG through the grant KL 1266/9-1 within the framework of the joint German-Russian project ``New constraints on nuclear parton distribution 
functions at small $x$ from dijet production in $\gamma A$ collisions at the LHC".

\end{document}